\documentclass[pre,aps,twocolumn,showpacs,superscriptaddress,floatfix]{revtex4}
\input epsf

\usepackage{amssymb}
\usepackage{graphicx}
\usepackage{amsmath}
\usepackage{latexsym}
\usepackage{verbatim}

\begin{document}

\title{Random field Ising model on networks with inhomogeneous connections}

\author{Sang Hoon Lee}
\email{lshlj@stat.kaist.ac.kr} \affiliation {Department of Physics,
Korea Advanced Institute of Science and Technology, Daejeon 305-701,
Republic of Korea}

\author{Hawoong Jeong}
\email{hjeong@kaist.ac.kr} \affiliation {Department of Physics,
Korea Advanced Institute of Science and Technology, Daejeon 305-701,
Republic of Korea}

\author{Jae Dong Noh}
\email{jdnoh@uos.ac.kr} \affiliation {Department of Physics,
University of Seoul, Seoul 130-743, Republic of Korea}

\date{\today}

\begin{abstract}
We study a zero-temperature phase transition in the random field
Ising model on scale-free networks with the degree exponent
$\gamma$. Using an analytic mean-field theory, we find that the
spins are always in the ordered phase for $\gamma<3$. On the other
hand, the spins undergo a phase transition from an ordered phase to
a disordered phase as the dispersion of the random fields increases
for $\gamma > 3$. The phase transition may be either continuous or
discontinuous depending on the shape of the random field
distribution. We derive the condition for the nature of the phase
transition. Numerical simulations are performed to confirm the
results.
\end{abstract}


\maketitle

\section{introduction}
Recently physicists have recognized that the underlying graphs or
networks for interacting systems have an intriguing structure. Such
complex networks are distinct from the periodic lattices in
Euclidean space in many aspects. The structural property of the
complex networks has been studied
intensively~\cite{Albert2002,Newman2003a,Dorogovtsev2001a}. Besides
the network topology itself, traditional topics of statistical
physics of complex networks have been investigated as
well~\cite{Pekalski2001,Hong2002,BJKim2001,Lopes2004,DJeong2005,
Sood2004a,Goltsev03,Dorogovtsev04,Igloi2002,Michard2005,SWSon2005}.
Statistical physical systems of networks are attractive since they
display theoretically interesting critical
phenomena~\cite{Goltsev03,Dorogovtsev04,Igloi2002}. They are also
attractive for a possible application to various phenomena in social
systems having a complex underlying network
structure~~\cite{Sood2004a,Michard2005,SWSon2005}.

In contrast to the periodic lattices in Euclidean space, complex
networks have an inhomogeneous structure. Recent studies reveal that
the structural inhomogeneity plays an important role in critical
phenomena on complex
networks~\cite{Sood2004a,Goltsev03,Dorogovtsev04,Igloi2002,Satorras2001,
Catanzaro2004}. However, in the study of the critical phenomena,
systems with quenched disorder have received little attention with
only a few exceptions~\cite{DHKim2005,Mooij2004}. In this work we
investigate the effect of quenched disorder and structural
inhomogeneity on the nature of a phase transition. For this purpose,
we study a disorder-driven phase transition in the random field
Ising model~(RFIM) on scale-free~(SF) networks. A SF network is
characterized by the power-law degree distribution $P_d (k) \sim
k^{-\gamma}$ with the degree exponent $\gamma$. The power-law
distribution indicates that SF networks have an inhomogeneous
structure and the degree exponent $\gamma$ determines the strength
of the inhomogeneity in structure.

The RFIM has attracted much attention in statistical
physics~\cite{Schneider1977,Aharony1978,Villain1984,Bruinsma1984,
Imbrie1984,Newman1996,Middleton2002,Swift1997,dAuriac1997}. Being
compared with the spin glass model where the quenched disorder is
present in the interaction among spins~\cite{MezardBook}, the RFIM
has a quenched random external magnetic field applied to each site.
The quenched disorder leads to a phase transition from an ordered
ferromagnetic phase to a disordered paramagnetic phase. In spite of
the simpler structure of the RFIM than the spin glass model, there
are still remaining questions and controversies, especially over the
nature of the phase
transition~\cite{Schneider1977,Aharony1978,Villain1984,Bruinsma1984,
Imbrie1984,Newman1996}.

We study the zero-temperature phase transition in the RFIM using an
analytic mean-field theory. Our analysis shows that the shape of the
random field distribution and the degree exponent $\gamma$ determine
the nature of the disorder-driven phase transition. We also perform
numerical simulations, which confirm the analytic results. This
paper is organized as follows: In Sec.~II, an analytic approach
based on mean-field theory and its prediction of the phase
transition nature are provided. Numerical simulations follow in
Sec.~III, and Sec.~IV summarize our work.

\section{analytic mean-field theory}

The Hamiltonian of the RFIM on a network is given by
\begin{equation}
\mathcal{H} = - J \sum_{i < j} a_{ij} s_i s_j - \sum_i h_i s_i ,
\label{Hamiltonian}
\end{equation}
where $s_i = \pm 1$ is the Ising spin variable of node $i = 1, 2,\ldots,
 N$, $J > 0$ is the ferromagnetic coupling strength between
neighboring spins, and $a_{ij}$ is the adjacency matrix element of
the network. The matrix element $a_{ij}$ takes the value of $1~(0)$
if two nodes $i$ and $j$ are (not) connected via a link. The degree
of a node $i$ is given by $k_i = \sum_j a_{ij}$. We will set $J=1$
hereafter for notational simplicity. Here the external magnetic
field $h_i$ is a quenched random variable, which is distributed
identically and independently according to a distribution function
$p(h)$. We only consider a symmetric distribution -- that is, $p(h)
= p(-h)$. It is convenient to write
\begin{equation}
p(h) = \frac{1}{\Delta} p_0 \Big( \frac{h}{\Delta} \Big),
\label{p_0:def}
\end{equation}
where $p_0 (x)$ is a normalized [$\int p_0 (x) dx = 1$] function
determining the shape of the distribution and $\Delta$ is a measure
of the disorder strength.

The ferromagnetic coupling $J$ favors the ordered state with all
spins up or down. The external magnetic field, however, tends to pin
each spin to a random direction. The competition between them may
lead to a phase transition, which will be investigated at zero
temperature using mean-field theory. In the framework of the
mean-field theory, each spin $s_i$ is assumed to be in equilibrium
under the effective magnetic field $\tilde{h_i} \equiv \sum_j a_{ij}
m_j + h_i$ where $m_j \equiv \langle s_j \rangle$ is the average
local magnetization at node $j$. Hence, the average local
magnetization should satisfy the coupled mean-field equation
\begin{equation}
m_i = \langle s_i \rangle = \tanh \left( \beta \sum_j a_{ij} m_j +
\beta h_i \right), \label{SC}
\end{equation}
where $\beta = 1/(k_B T )$ with the Boltzmann constant $k_B$ and
temperature $T$.

Instead of solving Eq.~(\ref{SC}) directly, we make a simplification
by assuming that the local magnetization depends only on the degree
and the magnetic field -- that is, $m_i = m(k_i,h_i)$. It could be
valid when the network has no internal structure and all nodes with
the same degree and the same random field are statistically
equivalent~\cite{Catanzaro2004}. Then $m(k,h)$ should satisfy
\begin{equation}
m(k,h) = \tanh \left[ \beta \sum_{k'} \int dh' p(h') m(k',h')k P_d
(k'|k) + \beta h \right] ,
\label{SC3}
\end{equation}
where $P_d (k' | k)$ is the conditional probability that a
neighborhood of a node with the degree $k$ has the degree $k'$.
The conditional probability $P_d (k'|k)$ measures a correlation
between degrees of adjacent nodes.
Although many real-world networks
display a nontrivial degree correlation~\cite{Newman2002a}, we focus
our attention on uncorrelated networks in this work for analytic tractability.
It will be interesting to study the effect of the degree correlation on
critical phenomena, which we leave for future work.
Without the correlation, the conditional probability
is given by
\begin{equation}
P_d (k'|k) = k' P_d (k') / \overline{k} ,
\label{Pkk'}
\end{equation}
with the mean degree $\overline{k}$~\cite{Catanzaro2004}.

Now we define the order parameter
\begin{equation}
m = \sum_k \frac{k P_d(k)}{\overline k } \int dh ~ p(h) m(k,h)
\label{m_definition}
\end{equation}
as the weighted average of the local magnetization.
Using Eqs.~(\ref{SC3}) and (\ref{Pkk'}), one finds that the order parameter
should satisfy
\begin{equation}
m = \int dk \frac{kP_d(k)}{\overline k } \int dh ~ p(h) \tanh
( \beta mk + \beta h )  \label{SC4}
\end{equation}
in the continuum limit. We are interested in the zero-temperature
limit where $\beta\to +\infty$. Using $p(h) = p(-h)$ and $p(h) =
p_0(h/\Delta)/\Delta$, we finally obtain the self-consistency~(SC)
equation for the order parameter at zero temperature given by
\begin{equation}
m = f(m) \equiv \int dk \frac{kP_d(k)}{\overline k }
G(km/\Delta) , \label{SC5}
\end{equation}
where
\begin{equation}
G(x) \equiv 2 \int_0^{x} dx'\ p_0(x') .
\label{G:def}
\end{equation}

The SC equation depends on the network inhomogeneity through
$P_d(k)$, the shape of the random field distribution through $G(x)$, and
the disorder strength $\Delta$.
SF networks have the power-law degree distribution. We use the
following explicit form for the degree distribution for further analysis:
\begin{equation}
P_d(k) = c k^{-\gamma}
\label{P(k)}
\end{equation}
for $k\geq k_0$. Here $k_0$ is a cutoff and $c=(\gamma-1)
k_0^{\gamma-1}$ is a normalization constant. The $l$th moment of the
degree, if it exists, will be denoted as $\overline{k^l}$. As for
the magnetic field distribution, we assume that $p_0(x)$ in
Eq.~(\ref{p_0:def}) is analytic at $x=0$~\cite{comment}. Then, the
function $G(x)$ in Eq.~(\ref{G:def}) can be expanded as
\begin{equation}
G(x) = \sum_{n=0}^\infty b_{2n+1} x^{2n+1} ,
\label{g(x)}
\end{equation}
where $b_1 = 2 p_0(0)$, $b_3 = p_0 ''(0)/3$, and so on. It has the
limiting behavior that $G(x \to 0) = 0$ and $G(x \to \infty ) = 1$.

\begin{figure}
\includegraphics[width=\columnwidth]{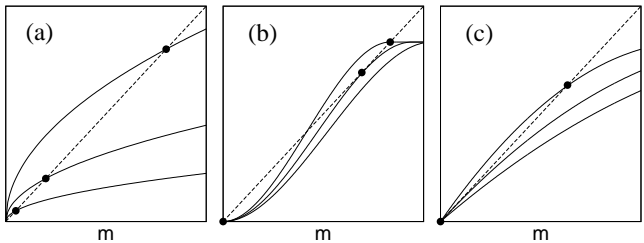}
\caption{Schematic plots of $f(m)$~(solid line), which has the
infinite slope at $m=0$ (a), is convex at $m=0$ (b) and is concave
at $m=0$ (c). The dashed line represents the graph of $m$, and the
solid circle represents the solution for the SC equation $m=f(m)$.}
\label{f(m)}
\end{figure}

The phase transition nature is determined by the leading behavior of
$f(m)$ near $m = 0$~(see Fig.~\ref{f(m)}). For a bounded degree
distribution -- e.g., Poisson distribution -- one can insert
Eq.~(\ref{g(x)}) into Eq.~(\ref{SC5}) and expand $f(m)$ into a
series of $m$ with odd-integer powers. However, with the power-law
degree distribution, the function $f(m)$ has a singular expansion.
In that case, one needs to split the function $G(x)$ into two parts
as $G(x) = G_r(x) + G_s(x)$ where $G_r(x) = \sum_{ n  <  \tilde{n} }
b_{2n+1} x^{2n+1}$ and $G_s(x) = G(x) -G_r(x)$. Here $\tilde{n}$ is
the largest integer among all satisfying $2n+1 < \gamma - 2$. We
will show that $G_r$ ($G_s$) contributes to $f(m)$ a regular
(singular) part consisting of integer (noninteger) powers of $m$.
Since the leading behavior of $f(m)$ depends on the value of
$\tilde{n}$, we consider the following three cases separately.

\subsection{$\gamma > 5$ case}
In this case, $\tilde{n}>1$ and $f(m)$ can be expanded as
\begin{equation}
\begin{split}
f(m) =& \sum_{n=0}^{\tilde{n}} C_{2n+1}
\left(\frac{m}{\Delta}\right)^{2n+1} +\\ & c
\left(\frac{m}{\Delta}\right)^{\gamma-2}\int_{mk_0/\Delta}^\infty dx
~ x^{1-\gamma} G_s(x),
\end{split}
\end{equation}
where $C_{2n+1} = b_{2n+1} \overline{k^{2n+2}}/\overline{k}$. Note that
$G_s(x) = \mathcal{O}(x^{2\tilde{n}+1})$ as $x\to 0$ and $G_s(x) =
\mathcal{O}(x^{2 \tilde{n}-1})$ as $x\to \infty$. This property
guarantees that the integral in the second term converges to a finite value.
Hence we find that
\begin{equation}
f(m) = \frac{ 2 p_0(0) \overline{k^2}}{\overline{k}} \left(
\frac{m}{\Delta}\right) + \frac{ p''_0(0) \overline{k^4}}{
3\overline{k}} \left(\frac{m}{\Delta}\right)^3 + \mathcal{O}
(m^\theta) ,
\label{SCE5}
\end{equation}
with $\theta = \min\{\gamma-2,5\}$.

The function $f(m)$ has a finite slope at $m=0$ and changes its
convexity depending on the shape of the random field distribution
given by $p_0(x)$. This property leads to the following conclusion:
For $p''_0(0)>0$, $f(m)$ is convex as shown in Fig.~\ref{f(m)}(b).
Then, the order parameter $m$ jumps from a nonzero value to zero at
a threshold value of $\Delta$. That is to say, the system undergoes
a {\em first-order phase transition}. For $p''_0(0)<0$, $f(m)$ is
concave as shown in Fig.~\ref{f(m)}(c). Therefore, the system
undergoes a {\em continuous phase transition} at $\Delta_c = 2 p_0
(0) \overline{k^2}/ \overline{k}$ and the order parameter scales as
\begin{equation}
m \sim ( \Delta_c - \Delta )^\beta ,
\label{m_scaling}
\end{equation}
with the order parameter exponent
\begin{equation}
\beta = 1/2  .
\label{beta5}
\end{equation}
These results coincide with those for the mean-field model where all
spins interact with all
others~\cite{Schneider1977,Aharony1978,Swift1997}.

\subsection{$3 < \gamma < 5$ case}
In this range of $\gamma$, $\tilde{n} = 1$ and $G_r(x) = b_1 x$. So
the function $f(m)$ is given by
$$
f(m) = \frac{b_1 \overline{ k^2} }{\overline{k}} \left(\frac{m}{\Delta}
\right)
+ c \left( \frac{m}{\Delta}\right)^{\gamma-2}
\int_{mk_0/\Delta}^\infty dx x^{1-\gamma} G_s(x).
$$
Note that $G_s(x)={\cal{O}}(x^3)$  as $x\to 0$ and $G_s(x) = {\cal{O}}(x)$
as $x\to \infty$ since $G_s(x) = G(x) - b_1 x$.
This property guarantees that the integral converges to a
finite value. So we find that
\begin{equation}
f(m) = \frac{2 p_0(0) \overline{k^2} }{ \overline{k}}
\left(\frac{m}{\Delta}\right) + cD
\left(\frac{m}{\Delta}\right)^{\gamma-2} + \mathcal{O}(m^3) ,
\end{equation}
where the constant $D$ is given by
\begin{equation}
D = \int dx ~ x^{1-\gamma} ( G (x) - 2p_0(0) x ) . \label{D:def}
\end{equation}

The function $f(m)$ has a finite slope at $m=0$ and changes its
convexity depending on the sign of the constant $D$. This leads to
the following conclusion: For positive $D$, the system undergoes a
{\em first-order phase transition}. For negative $D$, the system
undergoes a {\em continuous phase transition} at $\Delta_c =  2
p_0(0) \overline{k^2} / \overline{k}$ and the order parameter scales
as in Eq.~(\ref{m_scaling}) with the critical exponent
\begin{equation}
\beta = \frac{1}{\gamma-3} .
\label{beta35}
\end{equation}

We want to stress that the transition nature is determined by the
whole shape of the random field distribution given by the function
$p_0(x)$. For $\gamma>5$, it is determined by the sign of $p''_0(0)$
which is related to the local shape of $p_0(x)$ near $x=0$. On the
contrary, it is the sign of the constant $D$ that determines the
transition nature for $3<\gamma<5$. Hence, one may have the
continuous transition even with the magnetic field distribution with
$p''_0(0)>0$ and vice versa.

One may have a negative $D$ for a distribution $p(h)$ which has a peak at
$h=0$ and decreases monotonically as $|h|$ increases.
In such a case, the Ising spins become disordered
gradually as the disorder strength grows.
On the other hand, one may have a positive $D$
for a distribution $p(h)$ which has a peak at nonzero values of $h=\pm h_0$
and a deep valley at $h=0$. In such a case, the random
field breaks the order abruptly.

\subsection{$2 < \gamma < 3$ case}
In this range of $\gamma$, $\tilde{n} = 0$ and $G_s(x) = G(x)$. By
changing the integration variable $k$ to $x = mk/\Delta$ in
Eq.~(\ref{SC5}), and using Eq.~(\ref{P(k)}), we can write the
integral as
\begin{equation}
f(m) = c \left(\frac{m}{\Delta}\right)^{\gamma - 2}
             \int_{mk_0/\Delta}^\infty dx ~ x^{1-\gamma} G(x). \label{SC8}
\end{equation}
Note that $G(x)$ vanishes (at most) linearly as $x \to 0$ and
saturates to $1$ for $x \gg 1$. These properties guarantee that
the integral converges to a finite value in the limit $m\to 0$, which yields
that
\begin{equation}
f(m) = c' (m/\Delta)^{\gamma-2} +\mathcal{O}(m^1) ,
\end{equation}
with a constant $c'=c\int_0^\infty dx ~ x^{1-\gamma}G(x)$. The
function $f(m)$ has the infinite slope at $m=0$ and the SC equation
$m= f(m)$ has a nonzero solution
\begin{equation}
m \sim \Delta^{- (\gamma-2)/(3-\gamma)}
\label{m23}
\end{equation}
at all values of $\Delta$ [see Fig.~\ref{f(m)}(a)]. Therefore, the
system is ``always magnetized'' irrespective of the shape of the
random field distribution and the disorder strength. The SF network
with $2 < \gamma < 3$, where the second moment of degree diverges,
is famous for its peculiar behavior such as the absence of the
percolation and epidemic
threshold~\cite{Newman2003a,Satorras2001,Newman2001,Cohen2000}. This
``absence of magnetization threshold'' is another example of such
characteristic behavior.

In summary, we have a general criterion for the nature of the
zero-temperature phase transition of the RFIM with the symmetric
random field distribution $p(h)$ on SF networks with the degree
distribution $P_d (k) \sim k^{-\gamma}$ without the degree
correlation. For $2 < \gamma < 3$, the system is always magnetized
and there is no phase transition. For $\gamma > 3$, the system
displays a phase transition at a finite value of $\Delta$. The
transition may be either the first-order or continuous phase
transition. The condition for the first-order transition is that
$D>0$ [see Eq.~(\ref{D:def})] or $p''_0(0)>0$ for $3<\gamma<5$ or
$\gamma>5$, respectively. In the opposite case the transition is the
continuous one and the critical exponent for the order parameter is
given by Eq.~(\ref{beta5}) or (\ref{beta35}), respectively. Finally
we add a remark that a logarithmic correction appears when
$\gamma=3$ or 5.

\section{numerical simulation}
We perform a numerical study of the RFIM with the Hamiltonian in
Eq.~(\ref{Hamiltonian}) at zero temperature to confirm the analytic
result. First we generate a SF network of $N$ nodes and $K=2
\overline{k}N$ links  with the degree exponent $\gamma$ using the
so-called static model~\cite{KIGoh2001}. The static model has no
degree correlation except for the region where
$2<\gamma<3$~\cite{JLee2005}. Nevertheless, the ``always
magnetized'' characteristic of the system from the mean-field
analysis still holds for $2 < \gamma < 3$, as we will see. Random
magnetic fields are then assigned to each node according to a
distribution function $p(h) = p_0(h/\Delta)/\Delta$. The
ground-state spin configuration $\{s _i\}$ is found and the weighted
order parameter
$$ m = \frac{1}{\overline{k}N} \left| \sum_{i=1}^N k_i s_i \right| $$
is calculated, where $k_i$ is the degree of node $i$. The order
parameter is averaged over different samples to yield $\langle
m\rangle$. Note that each sample has a different realization of a
network configuration and a different realization of random fields.
The average over these samples corresponds to the order parameter
defined in Eq.~(\ref{m_definition}). The exact ground state of the
RFIM can be found numerically by adopting the mapping of the RFIM
onto the maximum flow problem~\cite{MaxFlowMinCut}. For details of
the mapping and the numerical algorithm solving the problem, we
refer reader to Ref.~\cite{MaxFlowMinCut}.

As for the random field distribution $p(h)=p_0(h/\Delta)/\Delta$,
we use the two functions $p_0(x) = p_+(x)$ and $p_0(x) = p_-(x)$ which
are given by
\begin{eqnarray}
p_+(x) &=& \frac{3}{2} x^2  \label{p_+}, \\
p_-(x) &=& \frac{\pi}{4} \cos \left(\frac{\pi x }{2}\right),
\label{p_-}
\end{eqnarray}
in the interval $-1\leq x \leq 1$ and zero outside the interval.
These functions have the following properties:
$p''_+(0)>0$ and $D>0$ for $p_+(x)$, and
$p''_-(0)<0$ and $D<0$ for $p_-(x)$.
So we can test the analytic result with these two distribution functions.

\subsection{Numerical result with $p(h) = p_+ (h/\Delta)/\Delta$}
\begin{figure}
\includegraphics*[width=\columnwidth]{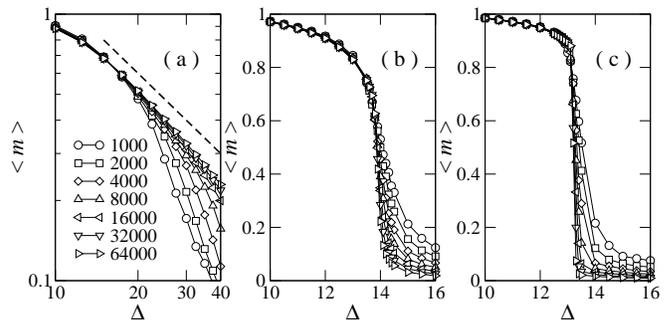}
\caption{$\langle m\rangle$ versus $\Delta$ with the magnetic field
distribution
$p(h) = p_+(h/\Delta)/\Delta$ and $\gamma=2.5$ (a), 4.0 (b), and 6.0 (c).
The dashed line in (a) has a slope $-1$.}
\label{fig:p+}
\end{figure}

We present the numerical data for the sample averaged magnetization
$\langle m\rangle$ in Fig.~\ref{fig:p+}. They were obtained from the
static model networks with $\gamma=2.5$, 4.0, and 6.0 of sizes
$N=1000,\ldots,64000$.

At $\gamma=2.5$, the ferromagnetic order with nonzero $m$ persists
at high values of $\Delta$. Moreover, the log-log plot in
Fig.~\ref{fig:p+}(a) suggests that the magnetization decreases
algebraically. This behavior is consistent with the analytic result
$m \sim \Delta^{-(\gamma-2)/(3-\gamma)}$ in Eq.~(\ref{m23}).
According to it, the decay exponent should be $-1$ at $\gamma=2.5$.
There is a little discrepancy in the decay exponent.  We attribute
the apparent discrepancy to a finite-size effect since the decay
exponent approaches the analytic result as $N$ increases. However,
we cannot exclude a possibility that it could be due to the negative
degree correlation at $\gamma=2.5$.

\begin{figure}
\includegraphics*[width=\columnwidth]{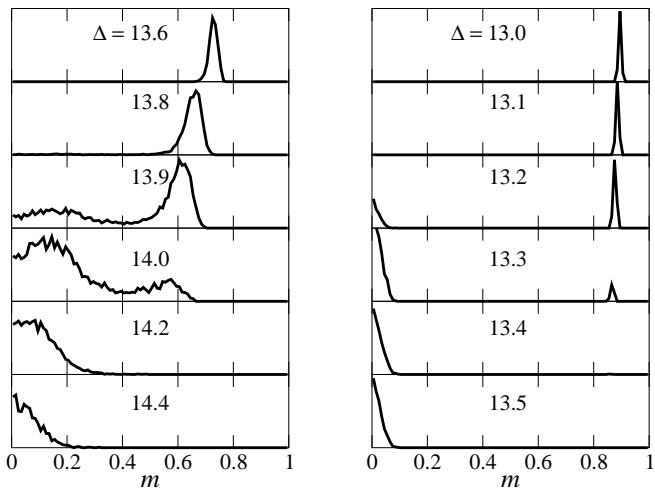}
\caption{The histogram on an arbitrary scale of the magnetization
at several values of $\Delta$
near the transition on the SF networks of $N=64000$ nodes
with $\gamma=4.0$~(left column) and 6.0~(right column).}
\label{fig:hist}
\end{figure}

Figures~\ref{fig:p+}(b) and \ref{fig:p+}(c) show that the order
parameter vanishes abruptly at a certain threshold of $\Delta$,
which is a characteristic of a first-order phase transition. In
order to prove the first-order nature we study the order parameter
histogram $H(m)$ near the threshold. Numerically the histogram is
measured by the fraction of samples whose order parameter value lies
between $m$ and $m+\delta m$, which is equal to $H(m)\delta m$. In
Fig.~\ref{fig:hist}, we present the histogram $H(m)$ obtained
numerically on the SF networks of $N=64000$ nodes with $\delta m =
0.01$. At small values of $\Delta$ the histogram is peaked at a
nonzero value of $m$, while it is peaked at $m=0$ at high values of
$\Delta$. In the intermediate values of $\Delta$, there appear two
peaks in the histogram, which indicates the phase coexistence. The
two-peak structure near the threshold confirms the first-order
transition nature.

\subsection{Numerical result with $p(h) = p_- (h/\Delta)/\Delta$}
\begin{figure}
\includegraphics[width=\columnwidth]{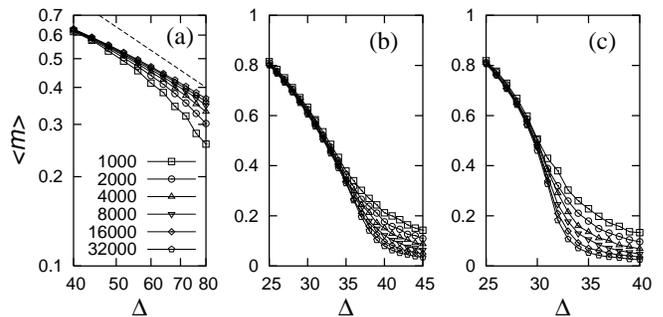}
\caption{$\langle m\rangle$ versus $\Delta$ with $p(h) =
p_-(h/\Delta)/\Delta$ and $\gamma=2.5$~(a), 4.0 (b), and 6.0 (c).
The dashed line in (a) has a slope $-1$.} \label{fig4}
\end{figure}

We present the numerical data obtained with the random field
distribution $p(h) = p_-(h/\Delta)/\Delta$ in Fig.~\ref{fig4}. At
$\gamma=2.5$ [Fig.~\ref{fig4}(a)], the order parameter remains
finite and decreases algebraically as $\Delta$, which is consistent
with Eq.~(\ref{m23}). At $\gamma=4.0$ and $6.0$ [Figs.~\ref{fig4}(b)
and \ref{fig4}(c)], the order parameter shows a threshold behavior.
Unlike the case with $p(h) = p_+(h/\Delta)/\Delta$, the order
parameter approaches zero smoothly as $\Delta$ increases. It
indicates that the transition could be a continuous transition.

In order to examine the transition nature, we measure the Binder
parameter~\cite{BindersCumulant}
\begin{equation}
U = 1 - \frac{ \langle m^4 \rangle }{3  \langle m^2
\rangle^2}. \label{Binder}
\end{equation}
The Binder parameter is supposed to take a nontrivial value
at a critical point with scale invariance.
It takes a trivial value $2/3$ and $0$ in an ordered phase and in
a disordered phase, respectively, in the $N\to \infty$ limit.
A critical point will manifest itself as a crossing point in the plot of $U$
versus $\Delta$ at different system sizes $N$.

\begin{figure}
\includegraphics[width=0.45\textwidth]{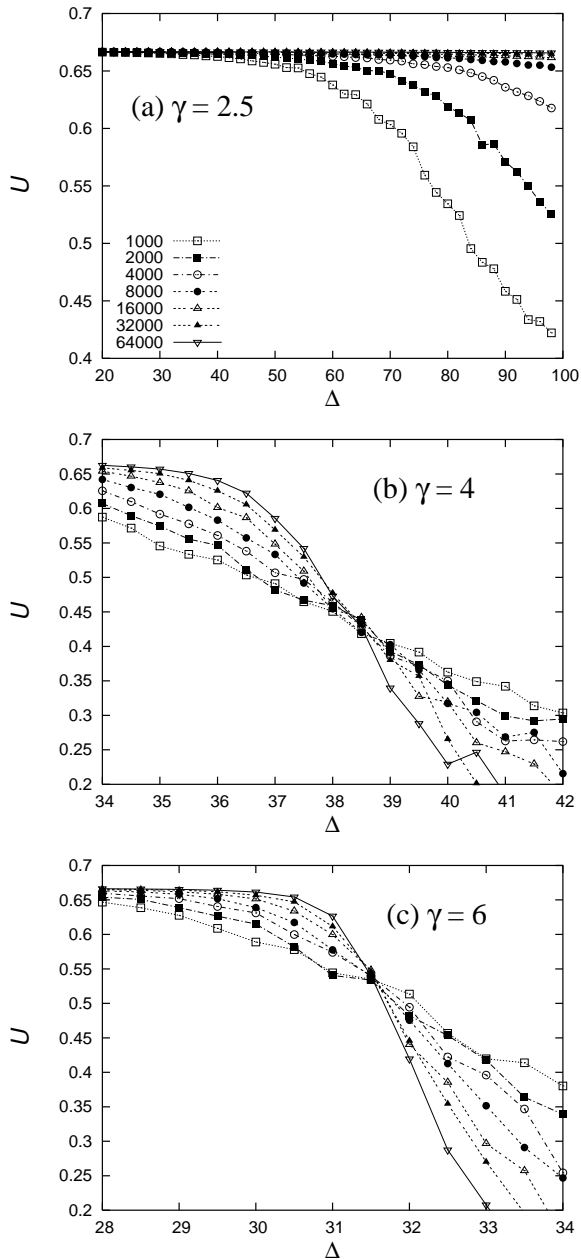}
\caption{$U$ versus $\Delta$ for various system sizes with
the degree exponent
(a) $\gamma = 2.5$, (b) $\gamma = 4$, and (c) $\gamma = 6$.}
\label{CosineBinder}
\end{figure}

Figure~\ref{CosineBinder} shows the Binder parameter for the three
cases, each of which corresponds to $2 < \gamma < 3$, $3 < \gamma <
5$, and $\gamma > 5$. In Fig.~\ref{CosineBinder}(a), there is no
crossing point and the value $U = 2/3$, corresponding to the ordered
state, persists as the system size grows. This behavior clearly
shows that the system is always magnetized in the thermodynamic
limit. Figures~\ref{CosineBinder}(b) and \ref{CosineBinder}(c) show
that there appear the crossing points at $\Delta_c\simeq$ $38.0$ for
$\gamma=4.0$ and $\Delta_c\simeq 31.5$ for $\gamma=6$ where the
Binder parameter is scale invariant and the system is critical. It
indicates that the transition is a continuous transition.

Since the phase transition is a continuous one, we expect that the
order parameter satisfies the critical finite-size-scaling
form~\cite{BindersCumulant}
\begin{equation}
\langle m \rangle = N^{-\beta/\nu'} F ( (\Delta_c-\Delta)^{\nu'} N )
, \label{fss}
\end{equation}
where $\beta$ is the order parameter exponent and $\nu'$ is the
finite-size-scaling exponent. The scaling function $F(x)$ has the
limiting behavior that $F(x\to 0) \sim \mbox{const}$ and $F(x\to
\infty) \sim x^{\beta/\nu'}$ so that
\begin{equation}
\langle m \rangle \sim (\Delta_c - \Delta)^\beta
\end{equation}
for $N \gg (\Delta_c-\Delta)^{-\nu'}$ and
\begin{equation}
\langle m \rangle \sim N^{-\beta/\nu'}
\end{equation}
for $N \ll (\Delta_c-\Delta)^{-\nu'}$.

\begin{figure}
\includegraphics[width=0.9\columnwidth]{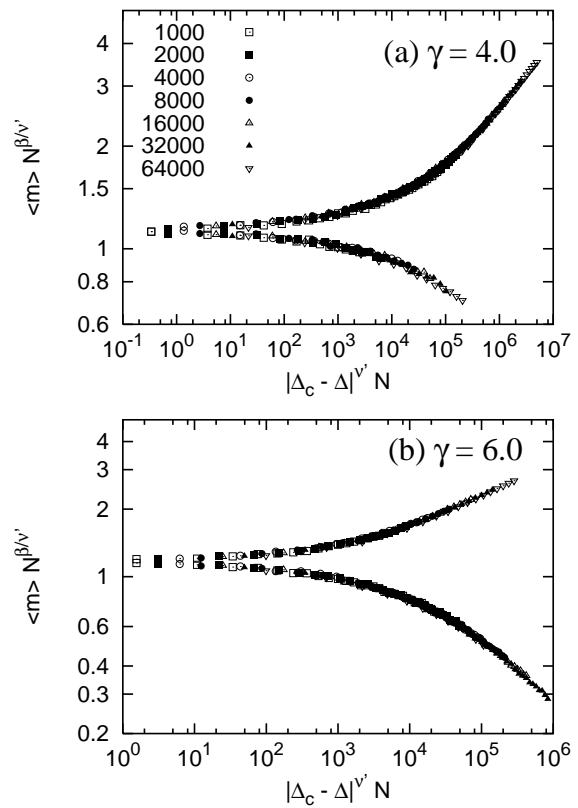}
\caption{Scaling plot of $\langle m \rangle N^{\beta/\nu'}$ versus
$|\Delta_c - \Delta|^{\nu'} N$ for $\gamma=4.0$ in (a) and
$\gamma=6.0$ in (b).} \label{fig6}
\end{figure}

The finite-size-scaling form is used to obtain the critical
exponents $\beta$ and $\nu'$. In Fig.~\ref{fig6}, we present the
scaling plot for the order parameter according to Eq.~(\ref{fss})
with the exponent values that give the best data collapse. We
estimate that $\beta = 0.75$ and $\nu' = 3.47$ for $\gamma=4.0$ and
$\beta = 0.45$ and $\nu'= 2.81$ for $\gamma=6.0$.

\begin{figure}
\includegraphics[width=0.9\columnwidth]{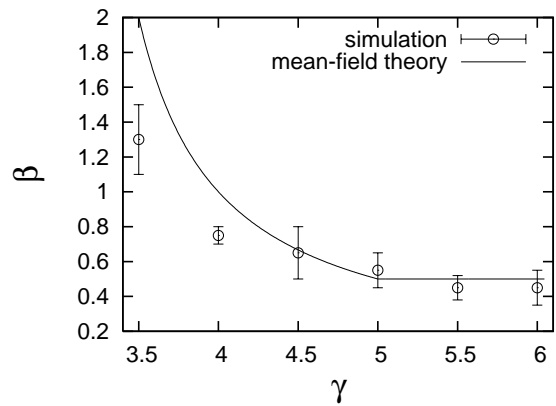}
\caption{The critical exponent $\beta$ as a function of the degree
exponent $\gamma$. The solid line corresponds to the mean-field
result $\beta = 1/(\gamma - 3)$ for $3 < \gamma < 5$ and $\beta =
1/2$ for $\gamma > 5$. The numerical simulation is based on the
finite-size-scaling method.} \label{BetaGamma}
\end{figure}

Repeating the same analysis, we obtained the values of $\beta$ at
several values of $\gamma>3$. The numerical results are compared
with the analytic mean-field results [see Eqs.~(\ref{beta35}) and
(\ref{beta5})] in Fig.~\ref{BetaGamma}. For $\gamma > 5$, the
critical exponent $\beta$ is indeed $1/2$ from the simulation
result. One finds that the values of $\beta$ for $3 < \gamma < 5$
deviate slightly from the mean-field prediction. Although we suspect
that this may be due to the singular dependence of
$\beta=1/(\gamma-3)$ near $\gamma=3$, we cannot exclude a
possibility that it may be due to a limitation of the our mean-field
approximation.

\section{discussion and conclusions}
The main result of our work is that the RFIM on scale-free networks
is always magnetized for $2 < \gamma < 3$ and it has a
disorder-driven phase transition for $\gamma>3$ whose nature depends
on the shape of the random field distribution
$p(h)=p_0(h/\Delta)/\Delta$. As for the nature of the transition,
similar results have been known in regular lattices in
high-dimensional Euclidean
space~\cite{Schneider1977,Aharony1978,Swift1997}: the
disorder-driven zero-temperature phase transition is first order or
continuous for a convex~($p''(0)>0$) or concave~($p''(0)<0$) random
field distribution, respectively. Our result shows that the same
criterion is valid for SF networks with $\gamma>5$. For
$3<\gamma<5$, the criterion is replaced by positivity or negativity
of the quantity $D$ defined in Eq.~(\ref{D:def}). Roughly speaking,
the distributions $p(h)$ highly peaked at $h=0$ give rise to the
continuous transition, while distributions highly peaked at nonzero
$h=\pm h_0$ give rise to the first-order transition. One may have a
distribution with $p''_0(0)>0$ but with $D<0$. For example,
$p_0(-1<x<1) = 3 [a + (1-a)x^2 ] / [2 (1+2a)]$ with $a = 3/4$ is
such a function. We checked numerically that it indeed leads to the
continuous phase transition at $\gamma=4$.

In summary, we have investigated the RFIM on scale-free networks
with inhomogeneous connections. The network topology, especially the
degree exponent $\gamma$, is shown to affect the phase transition
and the critical exponent. The shape of the random field
distribution is also responsible for the nature of the phase
transition.

\begin{acknowledgments}
This work was supported by Korea Research Foundation Grants Nos.
KRF-2004-041-C00139 (J.D.N.) and R14-2002-059-01000-0 (H.J.). The
authors thank Professor Doochul Kim and Professor Hyunggyu Park for
useful discussions.
\end{acknowledgments}


\begin{thebibliography}{99}
\bibitem{comment} Generalization to an arbitrary $p_0(x)$ is
straightforward.
\end{thebibliography}
\end{document}